# Effect of annealing on spinodally decomposed $Co_2CrAl$ grown via floating zone technique

*Ahmad Omar[‡]\*, Christian G. F. Blum[‡], Wolfgang Löser[‡], Bernd Büchner[‡§], Sabine Wurmehl[‡§]*

[‡]Leibniz Institute for Solid State and Materials Research IFW, D-01171 Dresden, Germany

[§]Institut für Festkörperphysik, Technische Universität Dresden, D-01062 Dresden, Germany

\*Corresponding author E-mail: a.omar@ifw-dresden.de

## Abstract

Among the large class of Heusler compounds, $Co_2CrAl$ is predicted to be 100 % spin polarized and is hence, a potential candidate for application in spintronics. So far, the predicted properties have not been experimentally realized which may be attributed to the phase segregated nature of samples. This phase segregation is avoided using floating zone growth. However, the grown sample was found to have undergone phase transformation via spinodal decomposition at low temperatures. In the present work, thermal annealing has been done on the spinodally decomposed samples and its effect on microstructure, crystallographic structure and magnetic properties has been studied. Annealing experiments were done and analyzed in order to understand the extent of the solid state miscibility gap. With regards to the phase diagram, the two-phase regime was found to extend till 1000 °C. Even at 1250 °C, we are still inside the immiscibility region. The thermodynamic miscibility gap was thus found to exist until high temperatures and alternate routes might be required to obtain a single phase sample with the desired high spin polarization and Curie temperature in the Co-Cr-Al system.

## Keywords:

## 1. Introduction

Heusler compounds are a very interesting class of compounds as they are known to display various appealing properties such as magnetoresistance, thermoelectric and shape memory effect etc. which have been nicely summarized in a comprehensive review [1]. Among the Co-based Heusler compounds, $Co_2CrAl$ has been predicted to be a half-metallic ferromagnet [2-5] and thrust has been in the past to advance this material for application in spintronics. Hence, $Co_2CrAl$ is also expected to show high magnetoresistance effect. Anomalous data have been reported in the past, both in bulk materials [6-9] as well as in films [10,11]. The anomalies may be attributed to chemical phase separation found in various samples [12-16].

Single phase samples become a necessity to understand the system and develop it from application point of view. Arc-melted samples of $Co_2CrAl$ are phase segregated, similar to an earlier work [6], as seen by the strong contrast in the backscattered electron (BSE) image (Fig.3a) although the two phases have similar structure and are thus indistinguishable in XRD. Recently, crystal growth using optical floating zone (FZ) technique was undertaken for the $Co_2CrAl$ compound in an attempt to grow single-phase sample as the FZ technique provides the possibility to grow incongruently melting systems. However, the grown sample was found to have undergone a phase transformation via spinodal decomposition [17]. In the present work, we study the effect of thermal annealing on the spinodally decomposed samples. We aim to gain a better understanding of the stability of the immiscibility gap in the system and thus make a step towards obtaining a single phase sample as well as understanding the phase transformations.

## 2. Experimental methods

Polycrystalline samples of the intended stoichiometry were prepared using a radio frequency (rf) induction melting Hukin-type copper cold-crucible device, in an argon atmosphere and cast to 6 mm diameter rods, 50–70 mm in length. Care was taken to avoid any oxygen contamination. The chamber was evacuated to $10^{-5}$ mbar pressure before backfilling with argon. The induction melting ensured homogeneous mixing of the constituents. The rods were used as seed and feed rods for crystal growth using the optical FZ technique. A vertical two-mirror Smart Floating Zone facility, designed and constructed at IFW Dresden [18,19] was used for the growth process. The growth experiment was carried out under constant argon flow of 0.15 l/min with additional gas cleaning by a Ti-getter furnace to minimize oxidation. The growth speed was 5 mm/h with counter-rotation of seed and feed rod of the order of ~0.3 Hz. During the growth, the temperature was measured using a pyrometer through a quartz window using a stroboscopic method [20]. The final zone was quenched in order to study the interface profile and zone composition. Regarding the annealing treatments, FZ-grown samples were sealed in evacuated quartz ampoules and annealed for 3 weeks at 400 °C, 750 °C, 1000 °C and 1250 °C respectively. All samples were quenched in water after annealing.

Samples were characterized using various techniques. Corresponding parts of the FZ-grown sample and annealed samples were cut and polished and analyzed using a NOVA-NANOSEM (FEI) high resolution scanning electron microscope (SEM) with an energy-dispersive X-ray (EDX) analyzer. Landscape image of the frozen zone was taken using Philips XL30 electron microscope. Structural characterization was done by X-ray diffraction (XRD) on powdered samples using a STOE STADI diffractometer in transmission geometry with Mo $K_{\alpha 1}$ radiation equipped with a Germanium monochromator and a DECTRIS MYTHEN 1K detector. For powder investigations, the samples were crushed by hand using a stainless steel mortar. The magnetic properties were investigated by a superconducting quantum interference device (SQUID, Quantum Design MPMS-XL-5).

## 3. Experimental results

### 3.1. FZ-grown sample

Fig.1a shows the FZ-grown rod of $Co_2CrAl$ marked with the beginning and the end zone. A small necking was given in the beginning to facilitate grain selection. Corresponding temperature measurement of the zone during the growth is shown in Fig.1b. The zone had stabilized after about 2 hours as the temperature fluctuations, albeit marginal, have settled down. The average temperature of the zone during most part of the stable growth was approximately 1540 °C. The temperature slightly dropped in the last hour likely due to evaporation and deposition on the quartz window of the growth chamber. However, no substantial effect of the evaporation was seen in the grown sample and the overall composition was nominal within the error of EDX measurement. The melting and crystallization interfaces are convex as seen in the landscape image of the frozen zone in Fig.1c. The microstructure of the final as-grown sample is an interconnected basket-weave pattern, very different from the dendritic two-phase microstructure of the as-cast sample (Fig.2a,b). The spatially ordered microstructure of the FZ-grown sample is attributed to a phase transformation via spinodal decomposition, the details of which have been discussed elsewhere [17]. The matrix phase is found to be cubic B2-type whereas the secondary phase is of tetragonal σ-CoCr type which is seen in the tetragonal splitting of the main reflection in the XRD data in Fig.3c. Rietveld refinement of the structural model was done. The lattice constant of the matrix was found to be 2.8672 Å and is consistent with TEM nanobeam diffraction experiments (not shown here, see Ref. [17] for details). The secondary phase ($P4_2/mnm$, a= b = 8.7650 Å, c = 4.5360 Å) has a phase fraction of 21.7 ± 0.5 vol %. For the details of the composition and refinement see Ref. [17].

## 3.2. Effect of thermal annealing

No distinguishable change, as compared to the FZ-grown sample, was observed in sample annealed at 400 °C. Whereas, the sample annealed at 750 °C exhibits a more pronounced decomposition which is apparent in the BSE image in Fig.3a. The secondary phase seems to be more interconnected. This is supported by the increase in the relative intensities of the σ-phase reflections in the XRD data in Fig.4.

The secondary phase is reduced in the sample annealed at 1000 °C although a small agglomeration in seen in the grains (Fig.3b). XRD data in Fig.4 shows a reduction in the relative intensities of the tetragonal splitting of the main reflection which is consistent with the lower phase fraction of the σ-phase. The sample annealed at 1250 °C is much different from the other samples. There is no strong contrast for the σ-CoCr phase in the BSE image (Fig.5c) although some variation in composition (chemical contrast) is surely visible. However, the boundaries between regions with different compositions are rather diffuse. Also, XRD does not show any tetragonal splitting of the main reflection and is pure B2-cubic as seen in Fig.4. The contrast seen in BSE image corresponds to two cubic B2-type phases with similar lattice constants, which are thus difficult to distinguish in XRD, similar to that observed in arc-melted samples [17]. These results of annealing suggest solid immiscibility within the B2-cubic $Co_2CrAl$ phase. Slight broadening of the B2 cubic reflections, compared to other annealed samples, is visible, in agreement with the presence of two phases with similar lattice constants.

## 3.3. Magnetic properties

All the samples show soft-magnetic behavior (*cf.* inset in Fig.5) at 5K. The saturation moments differ considerably from the Slater-Pauling value (3 µB/f.u.) and are summarized in Fig.5. The deviations in saturation moment are expected for both as-cast and FZ-grown (as-grown and annealed) $Co_2CrAl$, due to the two-phase microstructure. The magnetic moment of the as-grown FZ sample is considerably smaller than that of the arc-melted sample which is likely due to the well-developed secondary σ-phase. Saturation moment is further reduced after annealing at 750°C which can be attributed to the increased fraction of σ-phase, as seen in XRD and SEM. However, the reduction in the σ-phase fraction for sample annealed at 1000 °C, as seen in XRD and SEM, is also reflected in the magnetization as the saturation moment increases relatively. The saturation moment for sample annealed at 1250 °C is similar to that observed for the arc-melted sample, and further supports the fact that they are similar.

## 4. Discussion

Phase diagram data for the Co-Cr-Al system are limited and isothermal sections are only available for 900 °C and above [21-23]. Existence of a solid-state miscibility gap has only been demonstrated recently [17]. An adapted isothermal section at 1000 °C is also shown [17]. Based on the present study, it appears that, at 750 °C in the phase diagram, we are still inside the two-phase regime of the miscibility gap, as the extent of decomposition is increased due to annealing which is reflected in the relative intensities in XRD. Also, the saturation moment is further lowered which is understandable as the secondary σ-CoCr phase has been shown to have a strong effect on the magnetic properties in bulk [17] as well as in thin films [24]. After annealing at 1000 °C, the overall fraction of the secondary phase is less which is visible in the lower intensities for the tetragonal splitting of the main reflection as seen in Fig.4. Most likely, we are close to the phase boundary of the two-phase regime of the miscibility gap at 1000 °C. Due to the lower amount of the secondary phase, the saturation moment of the annealed sample has increased (1.47 µB/f.u.) compared to the FZ-grown sample (1.04 µB/f.u.) although it is still considerably below the Slater-Pauling value (3 µB/f.u.) as seen in Fig.5.

The sample annealed at 1250 °C is considerably different from the other annealed samples. The tetragonal secondary phase is dissolved completely and the sample exhibits two segregated cubic phases. The SEM, XRD and magnetization data confirm the structural and chemical similarity of the sample to that obtained through arc-melting. However, the formation route is completely different in both cases. The reflections are slightly broader in the case of FZ-grown sample annealed at 1250 °C, which is likely due to the diffuse nature of the phases. Regarding the phase diagram, even at 1250 °C, we may very well be in the immiscibility region of two cubic phases in the top part of the solid-state miscibility gap.

## 5. Summary & Conclusion

$Co_2CrAl$ was grown via FZ technique and was found have undergone a phase transformation via spinodal decomposition. FZ-grown samples were annealed at different temperatures in order to understand the stability of the solid-state miscibility gap. Based on our data, we can conclude that the solid state miscibility gap exists at least until 1250 °C in the phase diagram, although besides present data, so far nothing is reported in this regards in the literature. The two-phase regime of the miscibility gap in the phase diagram is up to the vicinity of 1000 °C. At 750 °C, we are well within the two-phase regime as we see a pronounced secondary phase in sample annealed at that temperature. It goes on to show the high stability of the σ phase and the tetragonal phase is even seen for annealed polycrystalline samples [6]. Thus, dissolving the secondary σ-CoCr phase via annealing, at least till 1250 °C, can be excluded as a possible way to obtain a single phase sample.

We have been able to get a better understanding of the immiscibility in the Co-Cr-Al system. Due to the large extent of the immiscibility, annealing does not allow to remove the secondary phase, at least till 1250 °C. One possible way may be, is to quench the sample during the FZ growth process. However, as the solid-state miscibility gap extends till high temperatures, quenching the sample is rather difficult, as an adequate thermal gradient is necessary at the crystallization front during the growth process. Therefore, alternate routes must be explored in order to overcome the issue and achieve a homogeneous phase-pure sample. Annealing treatments at temperatures between 1250 °C and the melting point at about 1500 °C, where one is possibly above the immiscibility region, is may be another possibility to obtain a single phase sample. However, this needs to be explored in the future experiments.

# Acknowledgements

S.W. gratefully acknowledges the financial support from Deutsche Forschungsgemeinschaft DFG in project WU595/3-1. The authors would like to thank Dr. L. Giebeler for assistance with the XRD measurement. The authors also thank Sabine Müller-Litvanyi and Gesine Kreutzer for support with metallography and zone images.

Figures

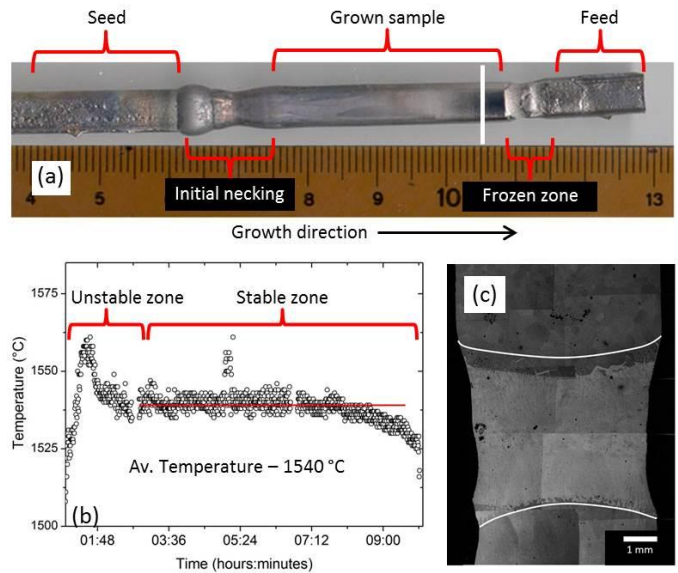

Figure 1: (a) FZ-grown Co$_2$CrAl sample showing the various parts of the grown rod. The samples for annealing were taken from around the region marked with the white line. (b) Pyrometer data for the growth showing periods of unstable and stable zone. (c) Frozen zone of Co$_2$CrAl growth showing convex interface (marked with white lines for clarity).

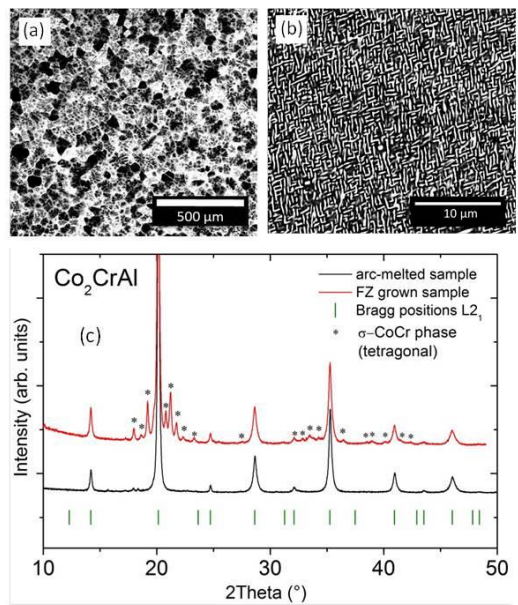

Figure 2: Co$_2$CrAl (a) SEM (BSE) image for as-cast sample (b) SEM (BSE) image for FZ-grown sample (c) XRD data for as-cast and FZ-grown samples where the reflections of the secondary σ-phase have been marked. The Bragg positions for ordered L2$_1$-type cubic structure are also shown.

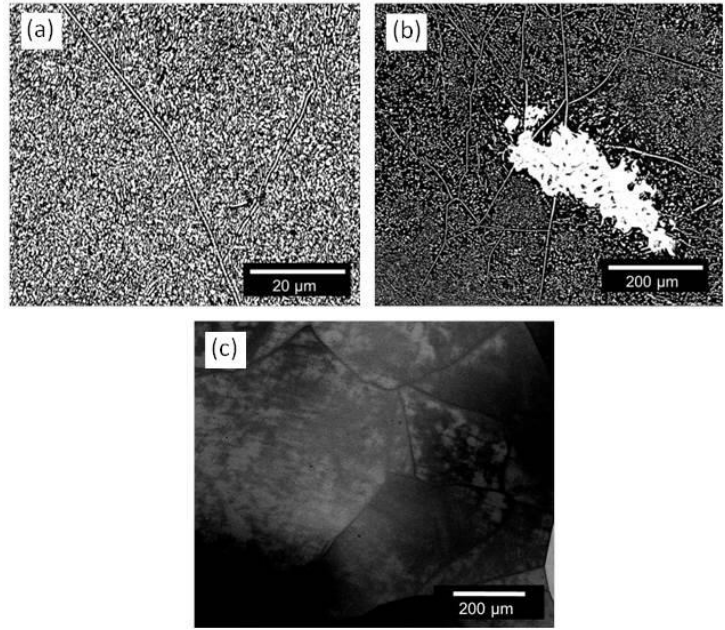

Figure 3: SEM (BSE) image for FZ-grown Co$_2$CrAl annealed for 3 weeks at (a) 750 °C, showing increased secondary σ-phase fraction (b) 1000 °C, showing lower secondary σ-phase fraction with agglomeration (c) 1250 °C, showing weak contrast between two cubic phases.

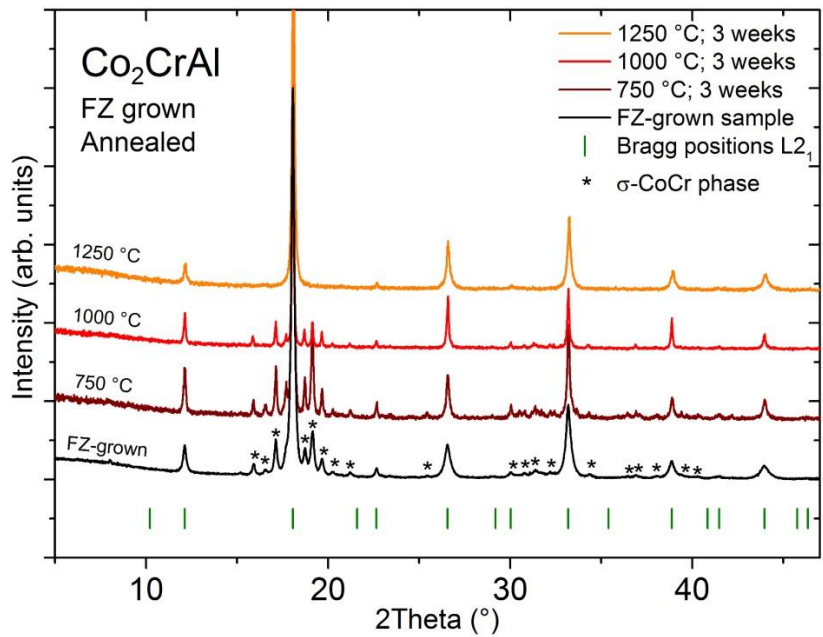

Figure 4: XRD data for annealed FZ-grown samples at different temperatures compared with the FZ-grown sample. The secondary σ-phase reflections have been marked. The Bragg positions for ordered L2$_1$-type cubic structure are also shown

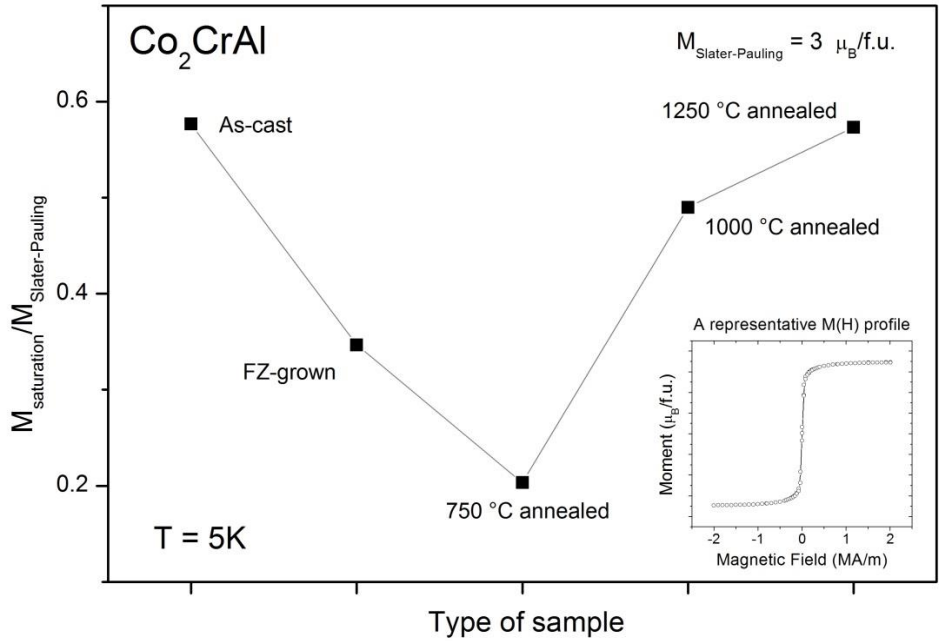

Figure 5: Relative saturation magnetization data for the various $Co_2CrAl$ samples. Annealed samples were kept at respective temperatures for 3 weeks and then quenched. Inset: M(H) curve for as-cast $Co_2CrAl$ sample.